\documentclass[aps,twocolumn,showpacs,amsmath,amsfonts,amssymb]{revtex4}
\usepackage{graphicx}
\usepackage{epsf}
\def\be{\begin{equation}}
\def\ee{\end{equation}}
\def\bea{\begin{eqnarray}}
\def\eea{\end{eqnarray}}

\def\a{\alpha}

\def\d{\delta}
\def\e{\epsilon}

\def\m{\mu}

\def\o{\omega}

\def\nn{\noindent}

\begin{document}
\title{\Large Sinusoidal excitations in reduced Maxwell-Duffing model}

\author{Utpal
Roy$^{\mathrm{1}}$, T. Soloman Raju$^{\mathrm{2}}$ and Prasanta K.
Panigrahi$^{\mathrm{1}}$\footnote{e-mail: prasanta@prl.res.in}}

\affiliation{$^{\mathrm{1}}$Physical Research Laboratory,
Navrangpura, Ahmedabad-380 009, India\\$^{\mathrm{2}}$Physics Group,
Birla Institute of Technology and Science-Pilani, Goa, 403 726,
India}


\begin{abstract}

Sinusoidal wave solutions are obtained for reduced Maxwell-Duffing
equations describing the wave propagation in a non-resonant atomic
medium. These continuous wave excitations exist when the medium is
initially polarized by an electric field. Other obtained solutions
include both mono-frequency and  cnoidal waves.

\end{abstract}
\maketitle

\noindent An atomic medium in general conditions is modelled by
N-level atoms. In the two-level resonant approximation, the system
is characterized by the Bloch equations, which is inaccurate in
several physical situations \cite{Cheng,Bowden}, like dense atomic
media and systems involving three or more level atoms. Thus the
resonant model needs to be generalized and extended to the
non-resonant scenario. One of the well studied approaches is to
consider the response of the medium as weakly nonlinear. Such
situation leads to the Duffing oscillator model, where the
nonlinear response of the medium is assumed to be cubic. This is
the simplest generalization of the Lorentz model.

On the other hand, Maxwell wave equation, for a linearly polarized
light, allows propagation in both the directions. However, this
can be approximated to unidirectional wave propagation when
anharmonic contribution to the polarization is very small. As a
result, the wave equation reduces  from second order to a first
order equation. For a non-resonant medium, this approximation
results in the reduced Maxwell-Duffing model (RMD).

Different excitations in non-resonant atomic media are currently
attracting considerable attention, because of their relevance to
ultra-short regime. Detailed reviews of various aspects of
non-linear excitations in atomic media can be found in
\cite{Bullough1,Bredec}. In case of two level atoms, the localized
soliton solutions of the Maxwell-Bloch equations explained the
physical phenomenon of self-induced transparency \cite{McCall}. In
the same system, general cnoidal waves have also been found as
exact solution \cite{Eberly1,Eberly2,Eberly3}. It was observed
\cite{Salamo} that, these waves can be naturally excited in the
presence of relaxation. Such shape preserving Jacobi elliptic
pulse train solutions have been experimentally observed
\cite{SalamoExp}. More general periodic solutions in multi-level
systems have also been reported \cite{PKPGSA}.

In case of Maxwell-Duffing model, a  class of exact localized
soliton solutions have been recently obtained \cite{Caputo}. We
present here mono frequency, sinusoidal wave excitations for RMD
system. This excitation exists only in the presence of a
polarizing background. General cnoidal wave solutions are found
both with and without background.

Below a brief summery of the reduced Maxwell-Duffing model is
presented after which a procedure to find exact solutions of this
system through a fractional linear transform is outlined. We then
present the novel sinusoidal wave solutions including single
frequency and cnoidal waves.

\begin{figure*}
\centering
\includegraphics[width=6.0 in]{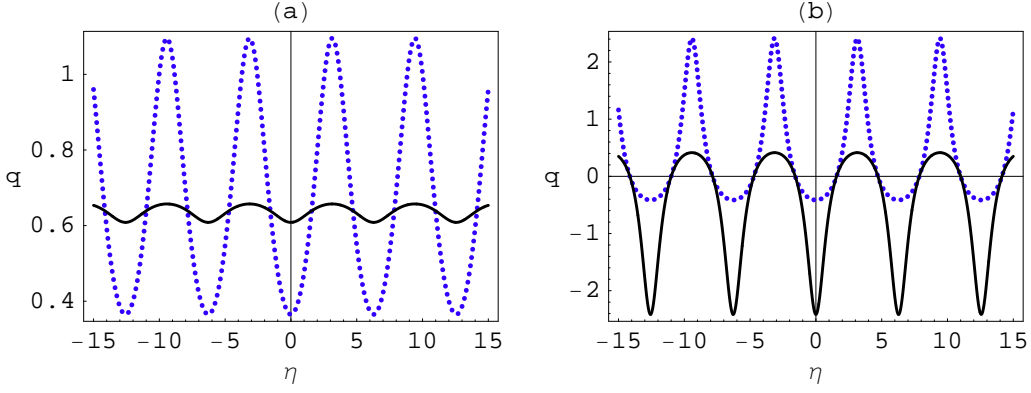}
\caption{Mono-frequency solutions for $\mu=0.5$ with (a)
$\alpha=1.6$ and (b) $\alpha=3.0$. The second case implies $A=0$.}
\label{2d}
\end{figure*}

\emph{Reduced Maxwell-Duffing Model:}

The propagation of electromagnetic waves in a medium is described
by the wave equation:
\begin{equation}
\frac{\partial^{2} E}{\partial
z^{2}}-\frac{1}{c^{2}}\frac{\partial^{2} E}{\partial
t^{2}}=\frac{4\pi}{c^{2}}\frac{\partial^{2} P}{\partial
t^{2}},\label{uniwave1}
\end{equation}
where $P$ is polarization of the medium. For unidirectional wave
propagation, the above equation can be reduced to a first order
equation,
\begin{equation}
\frac{\partial E}{\partial z}+\frac{1}{c}\frac{\partial E}{\partial
t}=-\frac{2\pi}{c}\frac{\partial P}{\partial t}.\label{uniwave2}
\end{equation}
In the Duffing oscillators model, the nonlinear response of the
medium is cubic. The corresponding equation for the motion of
electrons in the presence of an external electric field is given
by
\begin{equation}
\frac{\partial^{2} X}{\partial t^{2}}+\omega^{2}_{0}X+\kappa_{3}
X^{3}=\frac{e}{m}E. \label{duffing1}
\end{equation}
Here $X$ represents the displacement of an electron from its
equilibrium position, $\omega_{0}$ is the oscillator frequency,
$\kappa_{3}$ is anharmonicity constant, and $m$ is the effective
mass of the electron of charge $e$. The medium polarization is
defined as $P=n eX$, where $n$ is the number density of the
oscillators in the medium.

We choose new variables $\tau=z/l$, $x=\omega_{0}(t-z/c)$ and
normalize the independent variables as
\begin{equation}
 {\tilde e}=E/A_{0}, q=X/X_{0}.
\end{equation}
In terms of the new variables, Eq.~(\ref{duffing1}) then takes the
form
\begin{equation}
\frac{\partial^{2} q}{\partial x^{2}}+q+2\mu q^{3}={\tilde
e},\label{duffing}
\end{equation}
where $2\mu=\kappa_{3}X^{2}_{0}/\o^{2}_{0}$ and
$A_{0}=m\omega^{2}_{0}X_{0}/e=m\omega^{3}_{0}e^{-1}(2\mu/|\kappa_{3}|)^{1/2}$.
$X_0$ can be expressed as
$X_{0}=(2\mu\omega^{2}_{0}/|\kappa_{3}|)^{1/2}$. Similarly
Eq.~(\ref{uniwave2}) is transformed to
\begin{equation}
\frac{\partial {\tilde e}}{\partial \tau}=-\frac{\partial
q}{\partial x}.\label{wave}
\end{equation}
Here $l$ is defined as
\begin{equation}
l^{-1}=2\pi n e^{2}/(mc\o_{0})=\o^{2}_{p}/2c\o_{0},
\end{equation}
where $\omega_{p}=(4\pi n e^{2}/m)^{1/2}$ is the plasma frequency.
Eq.~(\ref{duffing}) and (\ref{wave}) together are called reduced
Maxwell-Duffing equations. For finding propagating solutions, one
defines
\begin{equation}
\eta=x-\tau/\alpha,
\end{equation}
where $\alpha$ is related to the velocity of the pulse.
Eq.~({\ref{wave}) can be integrated with respect to the single
variable $\eta$ to yield
\begin{equation}\label{efield}
{\tilde e}(t,x)=\alpha q(t,x)+ C,
\end{equation}
where $C$ is a constant, which signifies the background electric
field, when electron amplitude $q$ goes to zero. Non-linear equation
of motion in Eq.~(\ref{duffing}) then takes the form:
\begin{equation}
\frac{d^{2}q}{d\eta^{2}}+(1-\a)q+2\mu q^3=C\label{NLSEreal}.
\end{equation}
For a wave propagating in the right direction, $\a>1$ and $\m>0$,
which is considered below. The other propagation direction can be
like wise studied.

\emph{\textbf{Fractional Linear Transform and the solutions:}}

The above Eq.~(\ref{NLSEreal}) is in the form, which can be
obtained from the real part of the non-linear Schrodinger equation
with a source. For finding out explicit solutions we consider
Eq.~(\ref{NLSEreal}):
\begin{equation} \label{realeq}
q^{\prime\prime} +g q^3+{\e{ q}} = C,
\end{equation}
provided $g=2\mu$ and $\epsilon=(1-\alpha)$. Prime indicates
differentiation with respect to $\eta$. It is known earlier
\cite{solomon}, that this equation can be connected to the elliptic
equation $f^{\prime\prime}+a f+b f^3=0$ through the following
fractional linear transformation (FT):
\begin{equation}
\label{FT} q(\eta) = \frac {A + B f(\eta,m)^{\delta}} {1+ D
f(\eta,m)^{\delta}}
\end{equation}
\nn where $A$, $B$ and $D$ are real constants, $\delta$ is an
integer, and $f(\eta,m)$ is a Jacobi elliptic function, with the
modulus parameter $m$ ($0\leq m \leq 1$). It can be shown, that
$\d=2$ is the maximum allowed value. Here we concentrate on the
periodic solution of Eq.~(\ref{realeq}) for $\delta=1$ and $
q({\eta,m}) = cn({\eta},m) $. Solutions for other elliptic
functions also can be studied in a similar way.

\emph{I. General solution:}

i) The consistency conditions for $m=0$ are given by
\begin{widetext}
\begin{eqnarray}
\label{cond1}
A^{3}g+2D(AD-B)+A\epsilon-C=0,\\
\label{cond2}3A^{2}Bg+AD(1+2\epsilon)+B(\epsilon-1)-3C D=0,\\
\label{cond3}3AB^{2}g+AD^2(\epsilon-1)+BD(1+2\epsilon)-3C D^2=0,\\
\label{cond4}B^{3}g+BD^2 \epsilon-C D^{3}=0.
\end{eqnarray}
\end{widetext}

\begin{figure*}
\centering
\includegraphics[width=6 in]{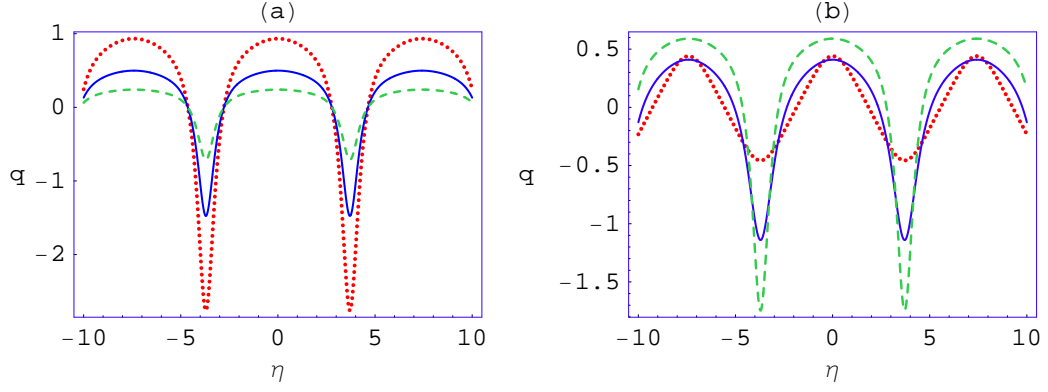}
\caption{Cnoidal wave solution in presence of source. (a) (Dotted
line) $g=2.0$, (solid line) $g=7.0$ and (dashed line) $g=30$ for
$\epsilon=-5.0$; b) (Dotted line) $\epsilon=-0.001$, (solid line)
$\epsilon=-2.0$ and (dashed line) $\epsilon=-5.0$ for
$g=5$.}\label{conoidal}
\end{figure*}

It should be pointed out that $cn({\eta},0)=cos{(\eta)}$. One can
see from the above FT (Eq.~(\ref{FT})) that $AD=B$ implies only a
constant or trivial solution and is not considered here. An $(AD-B)$
factor can be taken out of all the conditions by tactically using
the first consistency in Eq.~(\ref{cond1}). Other conditions were
used to solve for the unknowns $A$, $B$ and $D$. The source term
($C$) can be determined from the first condition itself. Thus, the
solution is expressed as
\begin{equation}
q(\eta) = \frac {A + B cos(\eta)} {1+ D cos(\eta)},
\end{equation}
where, $A$, $B$ and $D$ respectively are
\begin{eqnarray}\label{solution}
\nonumber &&A=\pm\frac{(\epsilon+2)}{\sqrt{3g(1-\epsilon)}},\; B = -
\sqrt{\frac{-(1+2 \epsilon)}{6g}},\\
&and&\;\;\;\;\;D=\pm \sqrt{\frac{-(1+2 \epsilon)}{2(1-\epsilon)}}.
\end{eqnarray}
After solving the solution parameters, the source term or the
constant electric field part can be determined from
Eq.~(\ref{cond1}):
\begin{equation}
C = -\frac{(1+2\epsilon)}{3}\sqrt{\frac{(1-\epsilon)}{3g}}
\end{equation}
As has been mentioned earlier, for wave propagation in the right
direction, $\mu>0$ and $\alpha>1$, implying $g>0$ and $\epsilon<0$
respectively. The parameter values exhibited in
Eq.~(\ref{solution}) yield the domain of the solutions: $g>0$ and
$\epsilon<-1/2$, where $A$, $B$, $D$ are real and $C$ is a
positive quantity. It is worth pointing out that all the solutions
are non-singular in nature. This is because the magnitude of $D$
is less than unity. The solutions are depicted in Fig.~\ref{2d}
(a) for $\mu=0.5$ with $\alpha=1.6$. The dotted line corresponds
to the solution for positive value of $D$ and the solid line is
for the negative one. The first plot is with a background,
\emph{i.e.}, $A\ne 0$.

ii) The consistency conditions, (\ref{cond1}-\ref{cond4}) support
solution for $A=0$ if the source is non-zero. In this case,
$\epsilon=-2$ and the solution is written as
\begin{equation}
q(\eta)=\pm \frac{1}{\sqrt{g}}\left(\frac{cos(\eta)}{\sqrt{2}\pm
cos(\eta)}\right),
\end{equation}
which is plotted in Fig.~\ref{2d} (b) for $\mu=0.5$ with
$\alpha=3.0$.

iii) Although a wide class of localized pulse propagation is
analyzed in \cite{Caputo} for different initial boundary
conditions, for completeness we indicate the same for $m=1$ in the
fractional transform. This is expressed as
\begin{equation}
q(\eta) = \frac {A + B sech(\eta)} {1+ D sech(\eta)},
\end{equation}
where
\begin{eqnarray}\nonumber
&A=\pm \sqrt{-\frac{(1+\epsilon)}{3g}}&\\\nonumber
 &D=\pm
\sqrt{-\frac{2(1+\epsilon)}{(1-2\epsilon)}},\;\;\; and \;\;\; B =
\pm \sqrt{\frac{2(\epsilon-2)^2}{3g(1-2\epsilon)}}&,
\end{eqnarray}
with $C^2 = -(1+\epsilon)(1-2\epsilon)^2/(27g)$. This solution
exists for $g>0$ and $\epsilon<-1$. For $\epsilon=2$, the solution
goes to the one, with $B=0$. All the solutions are non-singular
except for $B=0$ and $D=-\sqrt{2}$, which implies a singular one,
signifying self focussing effect.

iv) We now analyze the nature of the solutions in the absence of a
polarizing background: $C=0$. The resultant dynamical equation is
the real part of the well known non-linear Schr\"{o}dinger equation.
In this case, the periodic solution for the electron amplitude is of
quadratic fractional type ($\delta=2$): $A=\sqrt{-8/g}$, $B=0$,
$D=-2$ and $\epsilon=4$. The solution is singular one, implying an
instability in the electron amplitude or the self focussing of the
electric field. This happens for any value of $\mu<0$ and for
$\alpha=-3$.

v) In addition to the above mentioned mono-frequency solutions,
there are other conoidal wave solutions. As an example, when
$m=1/2$ the consistency conditions are
\begin{eqnarray}\nonumber
\label{cond11} A^{3}g+D(AD-B)+A\epsilon -C=0,\\\nonumber
\label{cond22}3A^{2}Bg+(2AD+B)\epsilon-3CD=0,\\\nonumber
\label{cond33}3AB^{2}g+D(AD+2B)\epsilon-3CD^2 =0,\\\nonumber
\label{cond44}B^{3}g+BD^2 \epsilon+(AD-B)-C D^{3}=0.\nonumber
\end{eqnarray}
These equations lead to,
\begin{eqnarray}\nonumber
&A=\sqrt{\frac{-E}{6g\epsilon}},\;\;\;B=-\frac{\epsilon}{g}\sqrt{\frac{g
F}{3 E}},\;\;D=\sqrt{\frac{-F}{2\epsilon}}&\\\nonumber
&and\;\;\;C=\frac{1}{3\sqrt{6}}\sqrt{-\frac{\epsilon}{g}}\left(\frac{4\epsilon^2
-F}{\sqrt{E}}\right),&
\end{eqnarray}
where $E=9+2\epsilon^2-3\sqrt{9+4\epsilon^2}$ and
$F=\sqrt{9+4\epsilon^2}-3$. $E$ and $F$ are positive for all
$\epsilon\ne 0$. These conoidal waves exist in the domain
$\epsilon<0$ and $g>0$. Solutions are picturised in
Fig.~\ref{conoidal}, where the variations of the electron
amplitude with $g$ and $\epsilon$ are displayed. As $\mu$
increases the amplitudes diminish for a fixed value of $\alpha=6$
(Fig.~\ref{conoidal}(a)). Fig.~\ref{conoidal}(b) shows the nature
of the solutions with $\alpha$ for certain value of $\mu=2.5$.

In conclusion, novel periodic solutions have been obtained for the
reduced Maxwell-Duffing model, which are unidirectional,
sinusoidal and of mono-frequency character. These solutions
complement the localized soliton solutions known in the literature
and occur only when a back ground electric field is present.
General cnoidal wave solutions are found both with and without
back ground fields. It will be interesting to study the nature of
the waves when Maxwell equation is modified by a non-local
dispersion term \cite{Chung}. Polarization properties of the
propagating modes should also be investigated in this non-linear
medium \cite{Song1}, as also the nature of the waves when slowly
varying amplitude approximation is not valid \cite{Leblond}.

\end{document}